\documentclass[aps,prd,twocolumn,groupedaddress,10pt,showpacs,nofootinbib]{revtex4} 
\usepackage{graphicx}% Include figure files
\usepackage{dcolumn}% Align table columns on decimal point
\usepackage{bm}% bold math
\usepackage{latexsym}
\usepackage{amsfonts}
\usepackage{amssymb}
\usepackage{amsmath}
\usepackage{color}
\begin{document}
\title{Ho\v{r}ava Gravity with Mixed Derivative Terms:\\ Power-Counting
Renormalizability with Lower Order Dispersions}
\author{Mattia Colombo,$^1$ A. Emir G\"umr\"uk\c{c}\"uo\u{g}lu,$^1$ and Thomas P. Sotiriou$^{1,2}$}
\affiliation{$^1$ School of Mathematical Sciences, University of Nottingham, University Park, Nottingham, NG7 2RD, UK\\
$^2$ School of Physics and Astronomy, University of Nottingham, University Park, Nottingham, NG7 2RD, UK} 
\date{\today}
\begin{abstract}
It has been argued that Ho\v rava gravity needs to be extended to include terms that mix spatial and time derivatives in order to avoid unacceptable violations of Lorentz invariance in the matter sector. In an earlier paper we have shown that including such mixed derivative terms generically leads to 4th instead of 6th order dispersion relations and this could be (na\"ively) interpreted as a threat to renormalizability.  We have also argued that power counting renormalizability is not actually compromised, but instead the simplest power counting renormalizable model is not unitary. In this paper we consider the Lifshitz scalar as a toy theory and we generalize our analysis to include higher order operators. We show that models which are power counting renormalizable and unitary do exist. Our results suggest the existence of a new class of theories that can be thought of as Ho\v{r}ava gravity with mixed derivative terms.
\end{abstract}
\maketitle
\section{Introduction}

The gravity theory proposed by Ho\v{r}ava in Ref.~\cite{Horava:2009uw} has acquired significant attention since its introduction. The basic idea is to improve the UV behaviour of the theory by modifying the dispersion relations, and hence the propagators. This is achieved by introducing a preferred foliation and constructing the action of the theory in such a way so that the `kinetic' terms contain only two time derivatives but there are also `potential terms' with higher order spatial derivatives. This introduces an anisotropic scaling between time and space at high energies, 
\begin{equation}\label{eq:scaling}
t\to [k]^{-m}t\,,\qquad
x^i\to [k]^{-1}x^i\,,
\end{equation}
where the latin indices span the $D$ dimensional spatial directions and $[k]$ is the momentum dimension.  It has been argued in Ref.~\cite{Horava:2009uw} that power counting renormalizability requires $m\ge D$, so in $3+1$ dimensions $m$ has to be at least equal to $3$ and the dispersion relations would be of the type $\omega^2\sim k^6$ in the UV.

 The existence of a preferred foliation leads to violations of Lorentz invariance in the gravity sector. One of the main challenges that the theory confronts with, is the percolation of Lorentz violations at low energy into the matter sector, where Lorentz symmetry is very stringently constrained (see e.g. \cite{Kostelecky:2008ts}). If dimension four Lorentz violating operators are present in the matter sector, the propagation speeds of different species of particles run to the universal value logarithmically, indicating a severe fine-tuning problem \cite{Collins:2004bp, Iengo:2009ix}. Even if such terms are absent (or tuned away) and the Lorentz violating operators are generated at higher dimensions, the latter are heavily constrained from synchrotron radiation in the Crab nebula \cite{Liberati:2012jf}. 

 A possible mechanism to suppress the Lorentz violations in the matter sector was proposed in \cite{Pospelov:2010mp}. Lorentz violations were restricted to the gravity sector at tree level and percolation to the matter sector  though graviton loops was considered. It was shown that the Lorentz-violating terms that are generated in the matter sector end up being suppressed by powers of $M_*/M_{p}$, where $M_*$ is the UV scale above which the dispersion relations in the gravity sector cease to be relativistic and $M_{p}$ is the Planck scale. Lorentz violation constraints in the gravity sector are quite weak as we do not test gravity at energies above $~10^{-2}{\rm eV}$. Hence, one can choose $M_*\ll M_{p}$ and this can push the Lorentz violations in the matter sector below the experimental constraints.

However, the analysis of Ref.~\cite{Pospelov:2010mp} also revealed a naturalness problem, stemming from the fact that the vector mode propagators are unaffected by the higher dimensional Ho\v{r}ava terms. As a result, the vector loops lead to quadratic divergences in the correction to the difference of propagation speeds between different matter species. The proposed resolution was to add the mixed derivative term $\nabla_i K_{jk}\nabla^i K^{jk}$ (see \cite{Pospelov:2010mp}\footnote{Curiously, mixed derivative terms also emerge in some Ho\v{r}ava--like extensions of supersymmetric models \cite{Gomes:2014tua}.}), where $K^{ij}$ is the extrinsic curvature of the leaves of the preferred foliation and $\nabla_i$ is the 3-dimensional covariant derivative operator on a leaf. Including this term in the action  improves the behavior of the vector mode.

This term is not the only operator with two temporal and two spatial derivatives that one could add. In Ref.~\cite{Colombo:2014lta} we considered all possible such terms and we performed the complete perturbative analysis of the most general extension of Ho\v{r}ava gravity along these lines. The dispersion relation of the scalar and tensor modes in the UV turned out to be of 4th order, i.e. $\omega^2 \propto k^4$, as opposed to the 6th order ones in standard Ho\v{r}ava gravity. One could interpret this as a threat to renormalizability, based on the standard power counting of Ho\v rava gravity. In fact, a specific tuning of coefficients that can restore the sixth order dispersion relations and that can still provide the sought for modification to the vector mode propagator does exist, so it is rather tempting to conclude that  this tuning is necessary. However, as shown in Ref.~\cite{Colombo:2014lta}, by studying the Lifshitz scalar as a toy model, counting rules get modified once the mixed-derivative terms are considered and one can have a renormalizable theory even with lower order dispersion relations. Surprisingly, the simplest power counting renormalizable theory of this type exhibits relativistic, instead of anisotropic, scaling and this, unfortunately, leads to problems with unitarity \cite{Colombo:2014lta}. 

In this paper, we revisit the Lifshitz scalar with mixed derivative terms as a proxy for the behavior of Ho\v rava gravity and show that, if a larger number of operators is taken into account, models that are power counting renormalizable and unitary exist and can have lower order dispersion relations than the standard Lifshitz scalar (or standard Ho\v rava gravity). 

\section{Lifshitz scalar with mixed derivative terms}\label{sec:dimcount}

In order to consider the mixed derivative case, we use, following Ref.~\cite{Colombo:2014lta}, the Lagrangian
\begin{equation}
{\cal L}=\alpha\,\dot{\phi}^2+\beta\,\dot{\phi}(-\triangle)^y \dot{\phi}-\gamma\,\phi(-\triangle)^z\phi\, .
\label{eq:lifshitzmixed}
\end{equation}
The anisotropic scaling is given in equation \eqref{eq:scaling}. The dimensions of the coupling constants are related through
\begin{equation}
[\alpha] = [\beta][k]^{2y}\,,\qquad
[\gamma]=[\beta][k]^{2(m+y-z)}\,.
\end{equation}
Hence, we can rewrite the Lagrangian (\ref{eq:lifshitzmixed}) as 
\begin{equation}
{\cal L} = \beta\left[\xi\,M^{2\,y}\dot{\phi}^2+\dot{\phi}(-\triangle)^y \dot{\phi}-M^{2\,(m+y-z)}\,\phi(-\triangle)^z\phi\right]\,,
\label{eq:lifshitzmixed2}
\end{equation}
where $[M]=[k]$ and $[\xi]=[k]^0$.

Here, we choose the normalization such that $\beta=1$ and fix the units such that the last two terms, which are expected to dominate in the UV, have the same dimensions. The latter condition gives the relation
\begin{equation}
m=z-y\,,
\label{eq:timescaling}
\end{equation}
which, given a theory with fixed $y$ and $z$, determines the degree of anisotropic scaling. 

Requiring that the action 
\begin{equation}
S = \int dt\,d^Dx\,{\cal L}\,,
\end{equation}
be dimensionless, the dimension of the Lifshitz scalar turns out to be
\begin{equation}
[\phi]= [k]^{d_{\phi}}=  [k]^{(D-m-2\,y)/2}\,.
\end{equation}
If the scalar field $\phi$ is dimensionless or has negative dimension, i.e. $d_\phi\le0$, the coupling constants of $\phi^n$ interactions with arbitrary positive integer $n$ has positive dimensions. The standard lore in quantum field theory dictates that positive dimensional coupling constants is an indication of renormalizability for the corresponding interactions. This translates into the condition \cite{Colombo:2014lta}
\begin{equation}
z = m+y \ge D-y\,.
\label{eq:renormfirst}
\end{equation}
Note that such dimensional arguments have to be treated with caution. Indeed, as we will show in Section \ref{sec:sdod}, they cease being trustworthy once derivative interactions are taken into account. We are interested in using the Lifshitz scalar as a proxy for understanding the UV properties of a gravity theory with the same anisotropic scaling properties and derivative structure. Since in gravity derivative interactions are inevitable, as a next step we include them as well and use a somewhat more robust criterion of renormalizability, the superficial degree of divergence.

\section{Superficial degree of divergence for derivative interactions}
\label{sec:sdod}

We now consider the free theory in the UV ($k \gg \xi^{1/2y} M$) by choosing the appropriate normalization and units in Eq.(\ref{eq:lifshitzmixed2})
\begin{equation} 
{\cal L}_{\rm UV} = \dot{\phi}(-\triangle)^y \dot{\phi}-\phi(-\triangle)^z\phi\,,
\end{equation}
The Green's function for the Lifshitz scalar can be calculated as
\begin{equation}
G_{\omega,k}=\frac{1}{\beta k^{2y}\,[\omega^2-k^{2\,m}]}\,.
\end{equation}
For the momentum cutoff $\Lambda$, each internal line in a Feynman diagram contributes 
\begin{equation}
G_{\omega,k} \to \Lambda^{-2(m+y)}=\Lambda^{-2z}
\label{eq:internal}
\end{equation}

Due to the anisotropic scaling the energy cutoff $\Lambda_\omega$ is different than the momentum cutoff $\Lambda$ and can be obtained through the dispersion relations as $\Lambda_\omega= \Lambda^m$. Thus, each loop integral contributes
\begin{equation}
\int d\omega d^Dk \to \Lambda_\omega\,\Lambda^D = \Lambda^{m+D}=\Lambda^{z+D-y}\,.
\label{eq:loop}
\end{equation}

We will consider the most general self-interaction term given by
\begin{equation}
{\cal L}_{\rm int} = \lambda\,(\nabla_i^{p_x}\,,\partial_t^{p_t}\,,\phi^s)\,,
\label{eq:generalinteraction}
\end{equation}
where $\lambda$ is the coupling constant, while $(\nabla_i^{p_x}\,,\partial_t^{p_t}\,,\phi^s)$ is shorthand for an $s$--particle operator that contains $p_x$ spatial derivatives, $p_t$ temporal derivatives, or $p\equiv p_x+m\,p_t$ weighted derivatives. The dimensions of the coupling constant can then be found as
\begin{equation}
[\lambda]=[k]^{d_{\lambda}} = [k]^{D+m-p-s\,d_\phi}\,.
\label{eq:lamdim}
\end{equation}
Assuming that all the derivatives in a given vertex arises from internal lines, the cutoff dependent contribution from each vertex will be $\Lambda^{p}=\Lambda^{p_x+m\,p_t}$. 

In conventional field theory, it is typically sufficient to have finite number of interactions that are renormalizable. However, here we are actually using a scalar field theory as a toy theory that will give us some insight into the renormalizability properties of Lorentz-violating gravity. The perturbative expansion of a  gravity theory includes infinitely many terms, due to the perturbative expansion of the inverse metric. All of these terms would have to be renormalizable for the theory to have the desired UV behaviour. Hence, what we need to require is that any interaction of the type (\ref{eq:generalinteraction}), with $s\to\infty$,  be renormalizable.  We purposefully avoid choosing any particular term from some specific theory as an example, as the renormalizability of any such term would not necessarily imply that the (nonlinear) gravitational analogue is renormalizable.

For a diagram with $L$ loops, $I$ internal lines, $E$ external lines and $V$ vertices, the superficial degree of divergence is calculated as\footnote{The assumption that all the momentum contributions at a given vertex comes from internal lines is a conservative one. Instead, if one imposes shift symmetry $\phi\to\phi+c$, all the external lines ($E$) would be associated with at least one spatial derivative of the field, contributing $-E$ to the right-hand side of (\ref{eq:generaldelta0}).} 
\begin{equation}
\delta \le L(D+m)-2I(m+y)+V\,p\,.
\label{eq:generaldelta0}
\end{equation}

Using two well-known identities, stemming from general properties of Feynman diagrams 
\begin{equation}
L-I+V=1\,,\qquad
s\,V=E+2\,I\,,
\label{eq:identities}
\end{equation}
we can extract more information from the superficial degree of divergence. To do so we first eliminate $L$ and $I$ using (\ref{eq:identities}) to find
\begin{eqnarray}
\delta \le D+m -d_{\phi}E - d_{\lambda}V\,,
\label{eq:generaldelta1}
\end{eqnarray}
where $d_{\phi}$ and $d_{\lambda}$ are the dimensions of the field and of the coupling constant, respectively.

This result is compatible with the standard intuition for power counting renormalizability: provided that $d_\phi >0$,
 any interaction with positive dimension coupling constant $d_\lambda>0$ will lead to a small and finite number or zero divergent diagrams, as convergence improves when the number of vertices or the number of external lines is increased. The condition $d_\lambda>0$ can then be interpreted as an upper bound on $s$ and $p$, see Eq.~(\ref{eq:lamdim}). 

When one wishes to use the Lifshitz scalar as a proxy for the behaviour of a gravity theory with the similar derivative structure, this standard result is not particularly useful. Gravity theories are highly nonlinear and linearization around a given background will generate an infinite number of terms with infinite copies of the field, albeit the limited number of derivatives in each term. Hence, one would wish to have convergent diagrams for any value of $s$. It is clear that this can only be achieved if $d_\phi\leq0$. 

Eq.~(\ref{eq:generaldelta1}) is not of great use when considering the $d_\phi<0$ case, as the external lines contribution comes with the wrong sign. However, using the identities in Eq.~(\ref{eq:identities}) one can rewrite Eq.~\eqref{eq:generaldelta0} as
\begin{equation}
\delta\le 2\, z +2\,d_{\phi}\, L - (2\,z-p)\,V\,.
\label{eq:generaldelta2}
\end{equation}
It is now straightforward to see that, so long as $d_\phi \le 0$,
the contribution from the loop either vanishes or each loop contributes more negative powers of the cutoff. It is  the number of vertices, or more specifically the number of weighted derivatives in a vertex that really determine how divergent the diagram is. 
For example, for nonderivative interactions $p=0$, we see immediately that the degree of divergence is $\delta\le 0$ if $d_\phi<0$, indicating that $\phi^n$ are either log divergent or finite \cite{Visser:2009fg}. 
For $0<p\le2z$ the vertices contributions to the degree of divergence are negative, making $\delta$ bounded from above by a finite value. In other words, for the interaction terms that have equal or less weighted derivatives than the free terms, there is a finite amount of counterterms that can remove the divergences. Interaction terms with $p>2z$ will be nonrenormalizable, as at a given loop order one can always have diagrams with an arbitrary number of vertices. Hence, such terms are not expected to be generated by radiative corrections.

To summarize, when derivative interactions are considered, in addition to (\ref{eq:renormfirst}), we obtain the second renormalizability condition which restricts the allowed number of derivatives in a given interaction
\begin{equation}
2\,z \ge p = p_x+m\,p_t\,.
\label{eq:renorm2}
\end{equation}
The maximum number of spatial gradients a renormalizable interaction can have is 
\begin{equation}
p_{x, {\rm max}}=2z\,,
\label{pmax}
\end{equation}
while the maximum number of time derivatives we can allow is
\begin{equation}
p_{t, {\rm max}}=\frac{2z}{m}=2+\frac{2y}{m}\,.
\label{rmax}
\end{equation}

We have thus found that the criterion for the renormalizability of an interaction term is more related to the number of derivatives it contains, rather than its dimensions. For the case where $d_\phi=0$, the two criteria do coincide as one can already see using Eq.~(\ref{eq:generaldelta1}); a term with a positive coupling constant necessarily contains equal or less derivatives than the free theory, thus is expected to be renormalizable. However, in the case of $d_\phi<0$, the intuitive description that links renormalizability with the dimensions of the coupling constant breaks down.  For instance, if $d_\phi$ is negative enough, $\nabla_i\phi$ can have negative dimensions and one can construct interaction terms with an arbitrary number of derivatives while still having a positive dimension coupling constant. Nonetheless, as we have shown above, the interaction terms with $p>2z$ would not actually be renormalizable.

\section{Restrictions from predictivity and unitarity}
\label{sec:unitarity}

The last point made in the previous section, regarding the fact that interaction terms with $p>2z$ are nonrenormalizable even though they have a positive dimension coupling constant, touches upon the issue of predictivity. If $d_\lambda>0$ were a sufficient condition for renormalizabilty for derivative interactions, then radiative corrections would generate infinite counterterms. A similar issue exists for interactions with $p<2z$ and a large number of copies of $\phi$: so long as $\phi$ has zero or negative dimensions, and for a given number of derivatives, there is an infinite number of renormalizable interaction terms with ever increasing copies of $\phi$ that do not carry derivatives. This has already been pointed out in \cite{Fujimori:2015wda} for the $y=0$ and $z=D$ theory, but our analysis reveals that it is actually a quite generic feature for  theories with $d_\phi\le0$. One need not worry about this problem for the Lifshitz scalar with no derivative interactions because it is a finite theory. But once derivative interactions are included the existence of infinite potential counterterms  poses an actual threat for predictivity. 
A simple solution is to impose some symmetry, e.g. a shift symmetry $\phi\to\phi+c$, thus rendering the number of terms finite \cite{Fujimori:2015wda}.  

In a gravity theory one expects to have such a symmetry anyway. In Ho\v rava gravity in particular, symmetry under foliation preserving diffeomorphisms (FDiffs) symmetry comes to the rescue. Although the expansion of the FDiff invariant terms lead to an infinite number of terms with ever increasing powers of the metric perturbations, the coefficients of these terms are not actually independent and can be expressed in terms of the original coupling constant, i.e. the number of coupling constants remain finite.

We can now turn our attention to unitarity. In a theory with derivative interactions one has to make sure that threatening terms such as $\ddot{\phi}^2$ will not be generated by radiative corrections. The second renormalizability condition (\ref{eq:renorm2}) implies that the total number of time derivatives a term can have is given by (\ref{rmax}), which can be larger than $2$ if $y>0$. In fact, the simplest example with $y=1$, $m=1$ studied in Ref.~\cite{Colombo:2014lta}  allows for dangerous terms with $4$ time derivatives and is thus nonunitary. As Eq.~(\ref{eq:timescaling}) indicates, the value of $m$ can be increased by including gradient terms with higher $z$ in the free theory. According to  Eq.~\eqref{rmax}, in order to avoid the unitarity breaking terms for $y>0$,  it is sufficient to require
\begin{equation}
m>y.
\label{unitarity1}
\end{equation}

Let us collect all of the conditions we have derived so far. For the theory
\begin{equation}
{\cal L}= \dot{\phi}(-\triangle)^y \dot{\phi}-\phi(-\triangle)^z\phi +\lambda\,(\nabla_i^{p_x}\,,\partial_t^{p_t}\,,\phi^s)\,,
\end{equation}
the power counting renormalizability and unitarity requirements lead to
\begin{eqnarray}
z&=&m+y\,,\nonumber\\
m&\ge& D-2\,y \,,\nonumber\\
2z &\ge& p_x + m\,p_t\,,\nonumber\\
m&>&y\,.\label{eq:condlist}
\end{eqnarray}
For $y=0$, $z=m$ and the standard renormalizability condition of Ho\v{r}ava gravity is recovered
\begin{equation}
z \ge D\,,
\end{equation}
along with the trivially satisfied condition $m >0$.

For $D=3$ and $y=1$, we obtain
\begin{equation}
z=m+1\,,\qquad
m > 1\,,
\end{equation}
where the last condition forbids relativistic scaling on the grounds of unitarity.

\begin{figure}
\includegraphics[width=7.5cm]{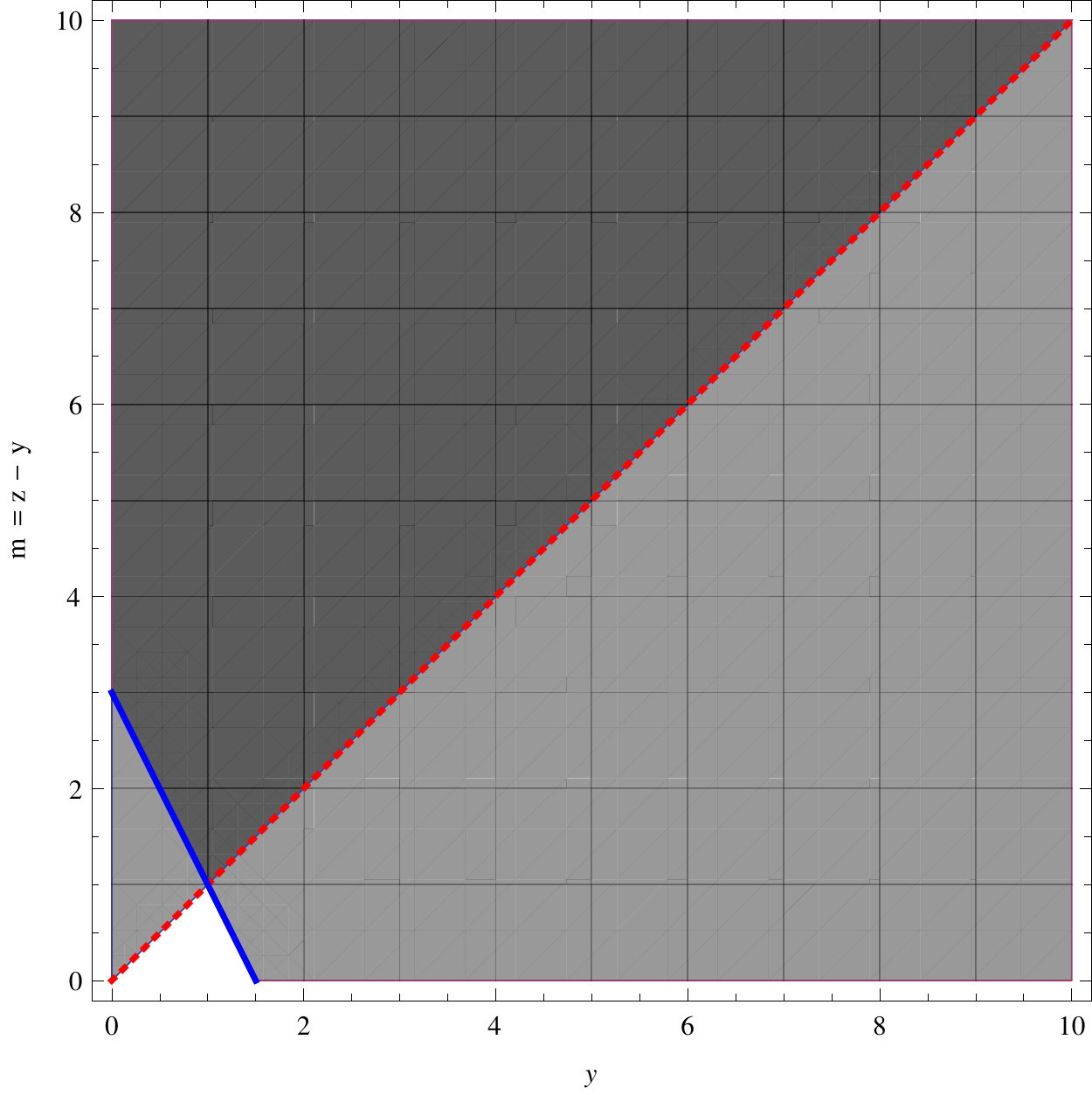}
\caption{For $D=3$, the allowed region for the scaling exponent and mixed derivatives. The region above (including) the solid blue line corresponds to the region where the renormalizability condition (\ref{eq:renormfirst}) holds. The region above (excluding) the dotted red line corresponds to the region where higher order time derivative terms are not generated (\ref{unitarity1}). The combined allowed region is the darkest region.
}
\label{fig:region}
\end{figure}

In Fig.~\ref{fig:region}, we show the allowed $(m,y)$ region in $D=3$. For a given mixed derivative term with arbitrary spatial derivatives, one can always satisfy the power counting renormalizability and the unitarity conditions, provided that enough powers of gradient terms are included in the free action.

\section{Discussion}\label{sec:concl}

To summarize, we have revisited the power counting arguments for a Lifshitz scalar with mixed derivative terms. We have gone beyond the analysis of Ref.~\cite{Colombo:2014lta} by considering the full class of theories with mixed derivative terms. Our results suggest that the addition of mixed derivative terms to the standard Lifshitz scalar can lead to theories that are power counting renormalizable and unitary, even though their dispersion relations and scaling properties are different than the standard Lifshitz scalar. We have identified the precise conditions, see Eq.~(\ref{eq:condlist}),  that define the subclass of theories with these characteristics. 

As discussed in the introduction, mixed derivative terms have been used in the context of Ho\v rava gravity in order  to regulate divergencies in vector mode loop integrals. These divergencies would otherwise introduce a naturalness problem in the suppression of Lorentz violations in the matter sector \cite{Pospelov:2010mp}. As shown in Ref.~\cite{Colombo:2014lta}, when all mixed derivative terms are consistently taken into account, the dispersion relations generically become 4th order. One could tune the coefficients in order to recover the 6th order dispersion relations (while still having the sought for modification to the vector mode propagator). However, to the extent that the Lifshitz scalar is a good proxy for Ho\v rava gravity, our results suggest that such tuning is not actually necessary. In $3+1$ dimensions, in particular a theory with $y=1$, $z=3$,  {\em i.e.}~standard Ho\v rava gravity with terms that are 6th order in spatial derivatives, supplemented with mixed derivative term with two temporal and two spatial derivatives, is both power counting renormalizable and unitary, even though it has anisotropic index $m=2$ and 4th order dispersion relations. 

We close with a note of caution: even though in all of the theories with mixed derivatives terms, the behavior of the vector mode gets modified, the dispersion relations for the scalar and tensor modes are not necessarily of sixth order. The analysis of Ref.~\cite{Pospelov:2010mp} regarding the suppression of Lorentz violations in the matter sector assumed sixth order dispersion relations, so it is not straightforward to conclude that its results would be applicable to theories with a different anisotropic index. One would have to revisit the problem in order to reach a final conclusion.

{\em Acknowledgments:} AEG and TPS would like to thank Shinji Mukohyama and IAP for their hospitality during the final stages of this work.
MC and TPS also thank the Perimeter Institute for its hospitality during the preparation of this manuscript. We are grateful to Kirill Krasnov, Jorma Louko and Matt Visser for illuminating discussions and helpful comments. 
The research leading to these results has received funding from the European Research Council under the European Union's Seventh Framework Programme (FP7/2007-2013) / ERC Grant Agreement n. 306425 ``Challenging General Relativity.''


\begin{thebibliography}{99}
%
\bibitem{Horava:2009uw} 
  P.~Horava,
  Phys.\ Rev.\ D {\bf 79}, 084008 (2009)
  [arXiv:0901.3775 [hep-th]].
%
\bibitem{Kostelecky:2008ts} 
  V.~A.~Kostelecky and N.~Russell,
  Rev.\ Mod.\ Phys.\  {\bf 83}, 11 (2011)
  [arXiv:0801.0287 [hep-ph]].
%
\bibitem{Collins:2004bp} 
  J.~Collins, A.~Perez, D.~Sudarsky, L.~Urrutia and H.~Vucetich,
  Phys.\ Rev.\ Lett.\  {\bf 93}, 191301 (2004)
  [gr-qc/0403053].
%
\bibitem{Iengo:2009ix} 
  R.~Iengo, J.~G.~Russo and M.~Serone,
  JHEP {\bf 0911}, 020 (2009)
  [arXiv:0906.3477 [hep-th]].
%
\bibitem{Liberati:2012jf} 
  S.~Liberati, L.~Maccione and T.~P.~Sotiriou,
  Phys.\ Rev.\ Lett.\  {\bf 109}, 151602 (2012)
  [arXiv:1207.0670 [gr-qc]].
%
\bibitem{Pospelov:2010mp} 
  M.~Pospelov and Y.~Shang,
  Phys.\ Rev.\ D {\bf 85}, 105001 (2012)
  [arXiv:1010.5249 [hep-th]].
%
\bibitem{Gomes:2014tua} 
  M.~Gomes, J.~R.~Nascimento, A.~Y.~Petrov and A.~J.~da Silva,
  Phys.\ Rev.\ D {\bf 90}, no. 12, 125022 (2014)
  [arXiv:1408.6499 [hep-th]].
%
\bibitem{Colombo:2014lta} 
  M.~Colombo, A.~E.~G\"umr\"uk\c{c}\"uo\u{g}lu and T.~P.~Sotiriou,
  Phys.\ Rev.\ D {\bf 91}, no. 4, 044021 (2015)
  [arXiv:1410.6360 [hep-th]].
%
\bibitem{Visser:2009fg} 
  M.~Visser,
  Phys.\ Rev.\ D {\bf 80}, 025011 (2009)
  [arXiv:0902.0590 [hep-th]].
%
\bibitem{Fujimori:2015wda}
  T.~Fujimori, T.~Inami, K.~Izumi and T.~Kitamura,
  arXiv:1502.01820 [hep-th].
%
\end{thebibliography}
\end{document}